\documentclass[epj,twocolumn]{webofc}
\usepackage[varg]{txfonts}   
\usepackage{graphicx}
\usepackage{subcaption}

\wocname{epj}
\woctitle{Seismology of the Sun and the Distant Stars 2016}
\begin{document}
\title{Galactic Archaeology with TESS: Prospects for Testing the Star Formation History in the Solar Neighbourhood}
%

\author{\firstname{Alexandra} \lastname{Thomas}\inst{1} \and
	\firstname{Emma} \lastname{Stevenson}\inst{1} \and
	\firstname{Fabian W. R.} \lastname{Gittins}\inst{1}\and
	\firstname{Andrea} \lastname{Miglio}\inst{1,2}\fnsep\thanks{\email{miglioa@bison.ph.bham.ac.uk}}\and
	\firstname{Guy} \lastname{Davies}\inst{1,2}\and
	\firstname{L\'eo} \lastname{Girardi}\inst{3}\and
	\firstname{Tiago L.} \lastname{Campante}\inst{1,2}\and
	\firstname{Mathew} \lastname{Schofield}\inst{1,2}
}

\institute{School of Physics and Astronomy, University of Birmingham, Edgbaston, Birmingham, B15 2TT, UK
	\and
	 Stellar Astrophysics Centre, Department of Physics and Astronomy, Aarhus University, Ny Munkegade 120, DK-8000 Aarhus C, Denmark
	\and
	Osservatorio Astronomico di Padova, Vicolo dell'Osservatorio 5, I-35122 Padova, Italy
}

\abstract{%
	
	A period of quenching between the formation of the thick and thin disks of the Milky Way has been recently proposed to explain the observed age-[$\alpha$/Fe] distribution of stars in the solar neighbourhood. However, robust constraints on stellar ages are currently available for only a limited number of stars.  The all-sky survey TESS (Transiting Exoplanet Survey Satellite) will observe the brightest stars in the sky and thus can be used to investigate the age distributions of stars in these components of the Galaxy via asteroseismology, where previously this has been difficult using other techniques. The aim of this preliminary study was to determine whether TESS will be able to provide evidence for quenching periods during the star formation history of the Milky Way. Using a population synthesis code, we produced populations based on various stellar formation history models and limited the analysis to red-giant-branch stars. We investigated the mass-Galactic-disk-height distributions, where stellar mass was used as an age proxy, to test for whether periods of quenching can be observed by TESS. We found that even with the addition of 15\% noise to the inferred masses, it will be possible for TESS to find evidence for/against quenching periods suggested in the literature (e.g. between 7 and 9 Gyr ago),  therefore providing stringent constraints on the formation and evolution of the Milky Way.	
}
\maketitle
%
\section{Introduction}
\label{intro}
\label{sec-1}
Uncertainties in the age distribution of stars in the solar neighbourhood have thus far limited the determination of the star formation history (SFH) of the local Galactic disk \cite{Astrophysics+and+Space+Science}, one of the key constraints to test competing models of formation and evolution of the Milky Way. It is hoped that this barrier may be overcome with the reception of all-sky asteroseismic data. 

In this study we investigated whether the TESS (Transiting Exoplanet Survey Satellite) mission \cite{Ricker2015} will be able to make a substantial contribution to the field of Galactic archaeology by providing evidence for quenching periods during the formation of the Galaxy. We used the TRILEGAL code (TRIdimensional modeL of thE GALaxy) \cite{TRILEGAL} to generate a set of synthetic stellar populations based on various SFH models, and considered only red-giant-branch (RGB) stars which would display oscillations detectable by TESS. 

The layout of this conference proceedings is as follows. We begin in Sect. \ref{sec-2} by describing the nature of the TRILEGAL population software and the code used to test simulated stars for detectable solar-like oscillations by TESS. In Sect. \ref{sec-3} we outline the SFH models used when generating  stellar populations. These are based around ideas of quenching of star formation rate (SFR) between formation of the thin and thick disks from \cite{Snaith14}. We then proceed in Sect. \ref{sec-4} to examine the results from analysis of age/mass distributions of the synthetic populations. We conclude in Sect. \ref{sec-5} with a discussion of whether the periods of quenching in the formation of the Milky Way explored in this study could be determined by analysing TESS data.

\section{Stellar Population Synthesis}
\label{sec-2}
The Monte Carlo population synthesis code, TRILEGAL  \cite{TRILEGAL}, was used to create all-sky populations of the Milky Way. The sky was split into 1280 tiles of equal areas for our first simulation and cylindrical symmetry was assumed. For subsequent populations, half of the sky was simulated for computational efficiency. For each tile, an extinction was calculated at infinity using a model provided by \cite{2003A&A...409..205D}. It was assumed that all stars could be considered to be resolved by TESS. This assumption would not necessarily be valid for those tiles in close proximity to the Galactic centre, and thus regions of the sky below 9 $\deg$ from the centre of the Galactic disk were omitted during analysis. 

Age is possibly one of the most distinguishing properties between the thin and thick disks. TRILEGAL models four components of the Galaxy separately: thin disk, thick disk, halo and bulge. Each component of the Galaxy is populated by stars generated from the same initial mass function (IMF) but with a different SFR and age-metallicity relation (AMR). For the purpose of this study, SFH of the bulge and halo were unchanged for all simulated populations. Models for the SFH of the two disk components are outlined in Sect. \ref{sec-3}.

Stars within the simulated populations were limited to red giants with detectable oscillations by TESS, as determined by the detection-probability algorithm described in  \cite{Campante16}. For this investigation we focused on RGB stars since they are expected to follow a well-defined age-mass relation, and so the ages of these stars can be determined thanks to asteroseismic constraints on their masses. 

\begin{figure}
	\centering
	\begin{subfigure}[b]{\linewidth}
		\includegraphics[width=\linewidth]{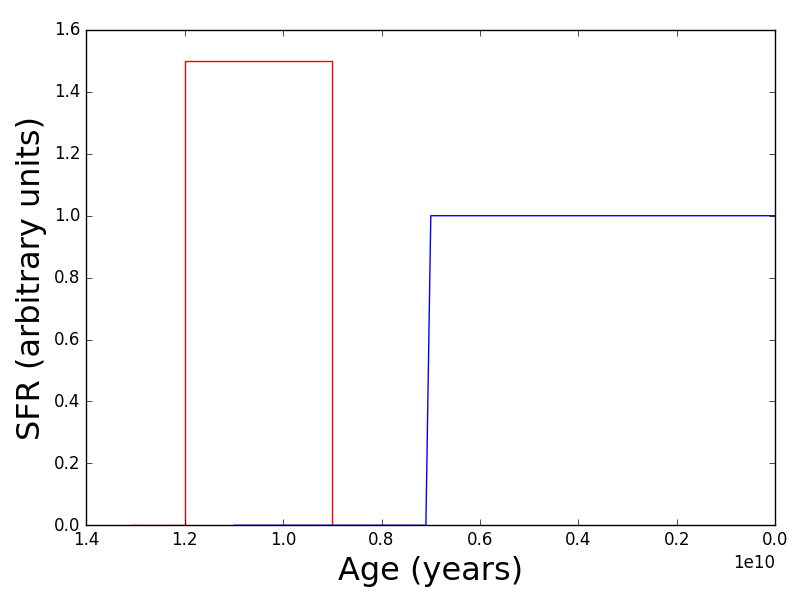}
		\caption{Model 1}
		\label{Fig: SFR2}
	\end{subfigure}
	\begin{subfigure}[b]{\linewidth}
		\includegraphics[width=\linewidth]{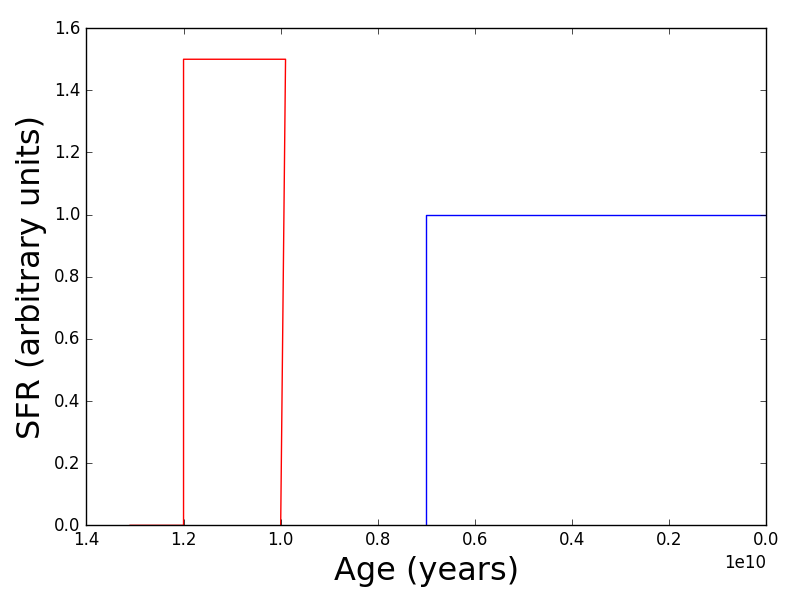}
		\caption{Model 2}
		\label{Fig: SFR4}
	\end{subfigure}
	\caption{Star-formation-rate models of the thick- and thin-disk components, used as a basis for 2 simulated populations using TRILEGAL. Thick-disk formation rate is represented in red, thin-disk SFR in blue.}
	\label{SFR models}
\end{figure}

\begin{figure}
	\centering
	\begin{subfigure}[b]{\linewidth}
		\includegraphics[width=\linewidth]{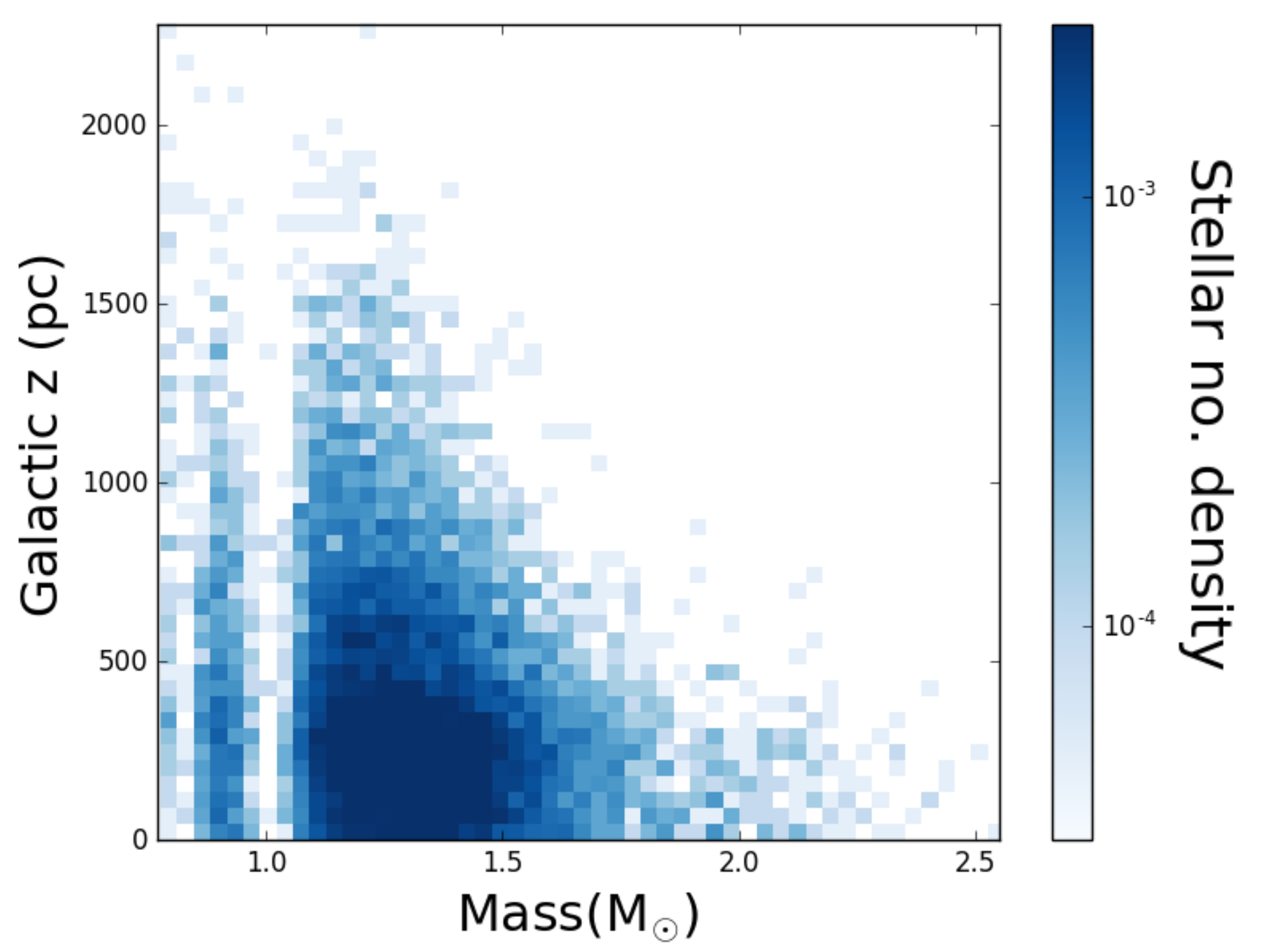}
		\caption{Model 1}
		\label{Fig: SFR2 age-mass}
	\end{subfigure}
	\begin{subfigure}[b]{\linewidth}
		\includegraphics[width=\linewidth]{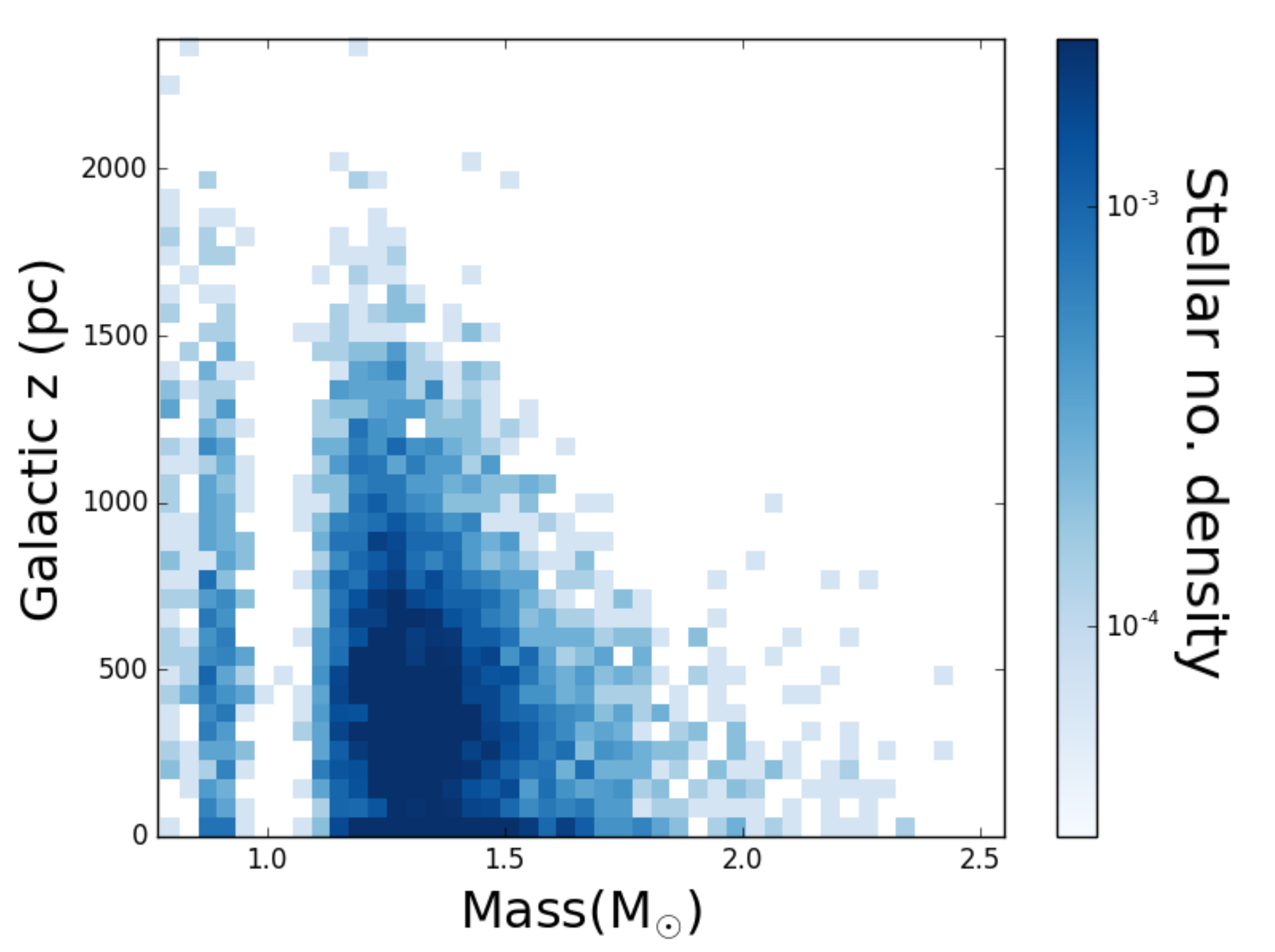}
		\caption{Model 2}
		\label{Fig: SFR4 age-mass}
	\end{subfigure}
	\caption{Two dimensional histogram distribution of RGB star mass (proxy for age) with disk height, $z$. Each distribution of synthesised stars was generated from the corresponding SFR models shown in Fig. \ref{SFR models}. }
	\label{Age-Mass distributions}
\end{figure}
\section{Star-Formation-Rate Models}
\label{sec-3}
The two disk components, which dominate the solar neighbourhood, were modelled to have distinct star-formation histories. Models were designed to test for the length of quenching period and durations of thin and thick disk formation.


Model 1 was based on ideas of quenching from \cite{Snaith14}, where a cessation in the SFR  between thick and thin disk formation is present. This mechanism can be used to explain the observed age-[$\alpha$/Fe] abundance distribution \cite{Snaith14} and is a phenomenon likely to depend on the mass of the galaxy \cite{2006A&A...453L..29C}. Using age ranges suggested by \cite{2016A&A...589A..66H}, \cite{2006MNRAS.369..673B},\cite{2011MNRAS.414.2893F} the thick-disk formation was chosen to extend from 12 to 9 Gyr ago and the thin disk from 7 Gyr ago to the present epoch. A quenching of duration 2 Gyr was used. The model also assumed a higher SFR during thick disk formation in comparison to the thin disk. This model can be seen in Fig \ref{Fig: SFR2}.

Model 2, shown in Fig. \ref{Fig: SFR4}, was based on the same ideas of quenching but used to study the consequences of a longer period of cessation since a range of quenching durations was found in literature \cite{2016A&A...589A..66H}. This model used a quenching gap of 3 Gyr extending from 10 to 7 Gyr ago.

\section{Results}
\label{sec-4}
\subsection{Age-Mass Distributions}
\label{sec-4.1}
In order to reconstruct the local star-formation history of the Milky Way, the age distributions of stars in the solar vicinity were studied. Since the stellar populations of the Galactic components are primarily discernible with height from the Galactic plane, $z$, we expect the age-$z$ distributions would differ for each SFR model. These are displayed in Fig. \ref{Age-Mass distributions} and are associated to the SFR models in Fig. \ref{SFR models}.

Red-giant-branch stars were chosen since these have a well-defined age proxy using the initial mass of the red giant's progenitor \cite[e.g. see][]{Stellar+structure+and+evolution} and so the ages of these stars can be determined by coupling asteroseismic constraints from TESS with photospheric properties from photometric/spectroscopic surveys \cite{Miglio2013, 2016arXiv160102802D}. 
Distances to stars may be determined either by seismic constraints or by dedicated astrometric missions, such as Gaia. 

From Fig. \ref{Age-Mass distributions} it can be seen that the distinct stages for formation of the thin and thick disk are separated by a thin gap in mass. It is obviously easier to discriminate models with longer quenching periods since the separation between these components is larger. 
For stars with $M\gtrsim 1.1\, {\rm M}_\odot$ the vertical structure of the mass distribution is determined by the assumption that the geometrical vertical scale height of the thin disk increases with age.

\subsection{Kolmogorov-Smirnov Testing}
\label{sec-4.2}
\begin{figure*}
	\centering
	\includegraphics[width=.7\linewidth]{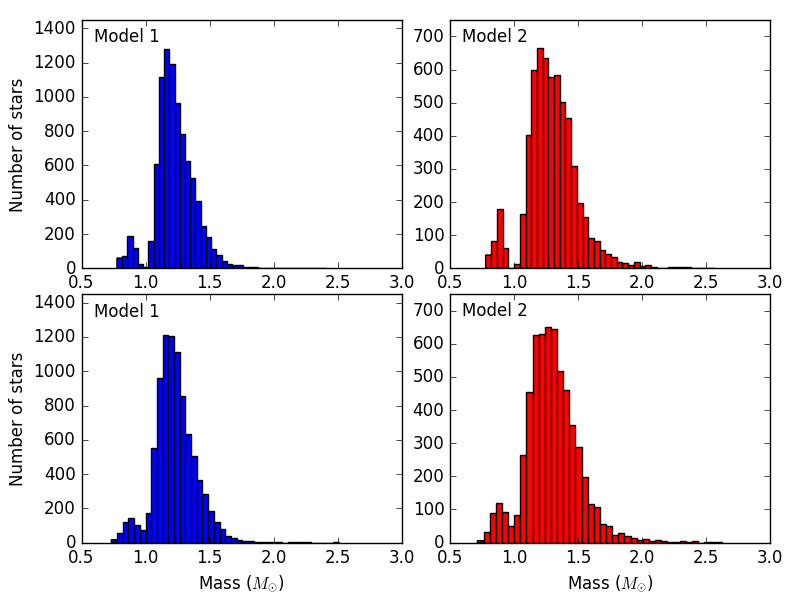}
	\caption{Histogram of mass distributions for models 1 (left) and 2 (right). Top row: No noise added. Bottom row: 15\% noise added.}
	\label{Fig: K-S test}
\end{figure*}

A Kolmogorov-Smirnov (K-S) test was conducted to determine whether TESS will be able to distinguish between populations modelled with varying duration of quenching period. Gaussian noise of 15\% \cite{2016arXiv160102802D} was added to the simulated mass data of models 1 and 2 by running 1000 Monte Carlo simulations (see Fig. \ref{Fig: K-S test}). The null hypothesis, that the data both with and without added noise came from the same distribution, was rejected. Models with different quenching lengths were found to be discernible, and only after 30\% noise was added were models no longer distinguishable. After the addition of 25\% noise model 2 showed quenching where model 1 did not. Since the expected error on asteroseismic mass is less than 30\% \cite{2013ARA&A..51..353C, 2016arXiv160102802D}, we can conclude that it should be possible for TESS to distinguish between models 1 and 2, and therefore to provide evidence for or against quenching. 

\section{Conclusions}
\label{sec-5}
The aim of this preliminary study was to investigate whether the all-sky asteroseismic mission TESS could significantly contribute to studies of Galactic archaeology in the solar neighbourhood. This was conducted by proposing two SFH models for the thin and thick disk which can be compared to data from TESS once observations have been made. The models used in this investigation tested for quenching periods during Galaxy formation and relative SFR of the disk components. 

The analysis was restricted to RGB stars since they have a well understood age-mass relation. Since the Sun is very close to the Galactic mid-plane, half sky populations were simulated.
Comparisons were made between distributions of stellar mass (proxy for age) with Galactic height for each model. A K-S test revealed that  with the addition of 15\% noise on the asteroseismically determined mass, we will be able to discriminate between star-formation quenching periods proposed in the literature. 

We have concluded that it will be possible for TESS to distinguish between models which include different lengths of quenching periods and therefore set an upper limit on the duration of a detectable quenching episode. The mission should be able to provide evidence for or against this scenario. It will be more difficult to determine relative SFR of the thin and thick disks, unless independent information (e.g. chemical abundances, kinematics) is considered. 

\vspace{2cm}
\begin{acknowledgement}
This work was done as part of a third-year undergraduate project at the School of Physics and Astronomy, University of Birmingham. AM, GRD, TC, and MS acknowledge the support of the UK Science and Technology Facilities Council (STFC). Funding for the Stellar Astrophysics Centre is provided by The Danish National Research Foundation (Grant agreement no.: DNRF106). LG acknowledges the support from the PRIN INAF 2014 -- CRA 1.05.01.94.05. 
\end{acknowledgement}

\bibliography{AlexandraThomas}

\end{document}